# The van der Waals force and gravitational force in matter

Lei Zhang*

School of Physics, Shandong University, Jinan, 250100, P. R. China

ABSTRACT

It was thought that the van der Waals force and gravitational force were distinct. Now a model is used to describe the attraction between macroscopic objects according to van der Waals interaction. The force between two objects with thermal equilibrium deviates from the law of universal gravitation slightly, and the gravity on the earth is explained approximately. We argue that the gravitational force is the van der Waals force actually. In other words, the gravitational force and mass are related to the quantum fluctuations of electron clouds in atoms, and these parameters are dictated by dielectric susceptibility.

* E-mail: Leizhang20@gmail.com

Gravitational force[1] and van der Waals force[2-12] are two different concepts. The van der Waals force originates from the instantaneous polarizability generated by the fluctuation of electron cloud in atoms.[2-12] It is the force that makes nonpolar atoms or molecules condense to liquids. It exists in air, liquid and solid universally. Its mechanism has been studied widely in nanoscience and liquid state theory. The gravitational interaction explains the motion of falling objects on the earth, the attraction among astronomical objects (such as earth and moon).[1] Its physical mechanism is still unclear.

Both of the van der Waals force and gravitational force are the attractive interactions among atoms. However, it was thought that they were effective in different scales. When the distance between two nanospheres is about several nms (Fig. 1A), the van der Waals force will exist between them, and the gravitational force is ignorable. When the distance between two nanospheres increases to 1000 km (Fig. 1B), the van der Waals force will disappear, and the gravitational force exists definitely in this scale. This tendency is particularly unusual. Persons will suppose that maybe the van der Waals force and gravitational force are equal to each other.

Some common physical characteristics exist in these two forces: (a). Both of them are attractive interactions among neutral atoms. (b). Both of them increase with the increase of the number of the electrons of neutral atoms. (c). Both of them increase with interacting body size. (d). Both of them exist universally everywhere in the world of matter.

The van der Waals force and gravitational force may follow different functions of distance. The gravitational force is given by[1]

$$F = GM_1M_2/D^2,  \qquad (1)$$

where $G = 6.67 \times 10^{-11} Nm^2/kg^2$ is the universal gravitational constant, $M_1$ and $M_2$ are the masses of the objects, $D$ is the distance between the centers of the objects. In nanoscale, the London–van der Waals interaction energy between two atoms A and B is given by[3]

$$E_{AB} = -1.5\alpha_A \alpha_B [I_A I_B /(I_A + I_B)]/D^6, \qquad (2)$$

where $\alpha_A$ and $\alpha_B$ are the dipole polarizabilities of the atoms, $I_A$ and $I_B$ are the characteristic energy of the atoms. The London dispersion force is $F_{AB} = 9\alpha_A \alpha_B [I_A I_B /(I_A + I_B)]/D^7$. When the $D$ is more than the wavelength corresponding to the atomic frequencies, the Casimir-Polder potential is[4]

$$E_{AB} = -23\hbar c \alpha_A \alpha_B /(4\pi D^7), \qquad (3)$$

where the $\hbar$ and $c$ are the reduced Planck constant and the speed of light. When the temperature $T \to \infty$, the potential in Lifshitz theory becomes[5,6]

$$E_{AB} = -3k_B T \alpha_A \alpha_B /D^6, \qquad (4)$$

where $k_B$ is Boltzmann's constant.

According to eqs.(2), (3) and (4), London force, Casimir force and Lifshitz force between objects of various geometries were derived.[4,5,7-9] The results explained many nanoscale phenomena. The retarded and unretarded van der Waals forces are nonadditive.[13] For the van der Waals force between two condensed objects, the correlation effect among closely packed atoms should be considered in any object. Recently, by multipole scattering techniques some results for Casimir forces between dielectric spheres or conducting spheres have been obtained.[9] At ground state, typically the potential at large separations between two compact objects 1 and 2 with arbitrary boundary conditions is[9]

$$E = \frac{\hbar c}{\pi} \sum_{j=3}^{\infty} \frac{B_j}{D^j} \approx -0.25 \frac{\hbar c}{\pi} \frac{C_1 C_2}{D^3}. \qquad (5)$$

The $C_1$ and $C_2$ are the capacitances of the objects 1 and 2. In macroscopic theory,[8] the retarded dispersion energy between two spheres 1 and 2 (Fig. 1D) is

$$E = -\frac{\hbar R_1 R_2}{\pi D} \int_0^\infty d\omega \operatorname{Re}\left(1 - \frac{\varepsilon_0}{\varepsilon_1}\right)\left(1 - \frac{\varepsilon_0}{\varepsilon_2}\right) \coth\frac{\hbar\omega}{2k_B T} X(k,D), \tag{6}$$

where $X(k,D) = \int_0^{2k} dq \cos qD \left[\frac{\sin qR_1}{qR_1} - \cos qR_1\right] \times \left[\frac{\sin qR_2}{qR_2} - \cos qR_2\right]\left(\frac{k^4}{q^4} - 0.5\frac{k^2}{q^2} + \frac{1}{8}\right)$. In eq.(6), the energy is a frequency integral ($\omega = ck$) for the dielectric constants $\varepsilon_1$ and $\varepsilon_2$ of the spheres 1 and 2. For small characteristic wave numbers $k$, eq.(6) omitting the temperature dependence becomes[8]

$$E = \frac{7\hbar c R_1 R_2 \mu}{24\pi D} \cdot \left\{\frac{1}{6}\left(D^2 + 3R_1^2 + 3R_2^2\right)\right.$$

$$\left. + \frac{1}{R_1 R_2}\left[\left[\frac{(D-R_1-R_2)^2}{24} - \frac{1}{6}\left(R_1^2 - R_1 R_2 + R_2^2\right)\right] \times (D+R_1+R_2)^2 \ln(D+R_1+R_2)\right]_{-R_1}^{R_1}\bigg|_{-R_2}^{R_2}\right\} \tag{7}$$

where $\mu$ is $\lim\limits_{k\to\infty}\left[k^4 \operatorname{Re}\left(1-\frac{\varepsilon_0}{\varepsilon_1}\right)\left(1-\frac{\varepsilon_0}{\varepsilon_2}\right)\right]$. Eq.(7) gives a finite value at zero separation ($d = D - R_1 - R_2$), and decreases with $D^{-3}$ at large $D$. It is consistent with eq.(5) for $D \to \infty$. For small characteristic wave numbers $k$ and $R_1, R_2 \ll D$, eq.(6) becomes

$$E = -\frac{\hbar c \mu R_1^3 R_2^3}{18\pi D} \int_0^\infty dk \coth\frac{\hbar\omega}{2k_B T} \int_{2k}^\infty dq\, e^{-qD}\left(1 - \frac{q^2}{2k^2} + \frac{q^4}{8k^4}\right) \tag{8}$$

The function $\coth(\hbar\omega/2k_B T)$ has an infinite number of poles, and equal to $\omega_n = i2\pi k_B T n/\hbar$. All $\omega_n$ with $n \geq 1$ are rather high for actual temperatures. Thus we keep only the term with n=0 in $\omega_n$. Now eq.(8) becomes

$$E = -\frac{\mu R_1^3 R_2^3 k_B T}{18 D^2} \tag{9}$$

For an atom, its atom polarizability $\alpha$ under an external field is $4\pi\varepsilon_0 a^3$ approximately. The $a$ is the radius of electron cloud (bound electrons) in the atom (Fig. 2A). For dielectrics (Fig. 2B), we have $\varepsilon' - \varepsilon_0 \propto N\alpha$, $N$ being the number density of atoms. For metals, their atoms include bound electrons (electron cloud) and free electrons. According to the equivalent circuit of metals (Fig. 2C), free electrons do not have a contribution for the ε′ when frequency f→∞. Thus we have $\varepsilon' - \varepsilon_0 \propto N\alpha$ for metals for f→∞.

The $\mu$ in eq.(9) is related to the dielectric susceptibilities of the spheres 1 and 2. Thus we have $\mu \propto N_1(4\pi\varepsilon_0 a_1^3) N_2(4\pi\varepsilon_0 a_2^3)$. The energy E is

$$E = -\eta \frac{N_1(4\pi\varepsilon_0 a_1^3)(4\pi/3)R_1^3 N_2(4\pi\varepsilon_0 a_2^3)(4\pi/3)R_2^3 k_B T}{D^2} = -\eta \frac{p_1 V_1 p_2 V_2 k_B T}{D^2} \qquad (10)$$

where the $\eta$ is a constant, the $p_1$ and $p_2$ are $N_1(4\pi\varepsilon_0 a_1^3)$ and $N_2(4\pi\varepsilon_0 a_2^3)$, the $V_1$ and $V_2$ are the volumes of spheres 1 and 2. The force F between the spheres 1 and 2 is

$$F = 2\eta \frac{p_1 V_1 p_2 V_2 k_B T}{D^3} \qquad (11)$$

In Fig. 1C, the core in the earth is composed of Fe with a small fraction of S and O, and the mantle is a mixture of silicates and a small fraction of metals.[14-16] Fig. 2D and E show their equivalent circuits. Similar circuits have been used to explain the dielectric behavior of polymer/metal composites.[17] According to the circuits in Fig. 2D and E, the free electrons in the core and mantle do not have a contribution for the ε′ when f→∞. The earth is a composite globe composed of numerous dielectric and conducting parts. Consequently, the interaction of a falling sphere and any small part of the earth can be described by eq.(5), (7) and (10).

The density of the earth increases from 2.6 to $13\times10^3$kg/m$^3$ when the position changes from surface to center.[16] Thus we suppose that the polarizability $p$ of earth is described by $p(r) \propto \rho_c - (\rho_c - \rho_s)r/R$. The $\rho_c$ and $\rho_s$ are the densities at the center and surface. The R is the radius of the earth, and the $r$ is the distance between the local position and the center of the earth. The temperature of the earth increases from 300K to 5600K when the position changes from surface to center.[14,15] It is supposed that the temperature of earth is described by $T(r) = T_c - (T_c - T_s)r/R$. The $T_c$ and $T_s$ are the temperatures at the center and surface. According to eq.(10), we obtain the energy E between the earth and a falling sphere by integration.

$$E \propto -(\rho_1 V_1) \left\{ \rho_c T_c \left[ \frac{r^2 - R^2}{2r} \ln\left(\frac{r-R}{r+R}\right) + R \right] - [T_c(\rho_c - \rho_s) + \rho_c(T_c - T_s)] \cdot \left[ \frac{r^2}{3R} \ln((r-R)(r+R)) \right. \right.$$

$$\left. \left. - \frac{R^2}{3r} \ln\left(\frac{r-R}{r+R}\right) + \frac{R}{3} - \frac{r^2}{3R} \ln r^2 \right] + (\rho_c - \rho_s)(T_c - T_s) \left[ \frac{r^4 - R^4}{4R^2 r} \ln\left(\frac{r-R}{r+R}\right) + \frac{r^2}{2R} + \frac{R}{6} \right] \right\} \quad (12)$$

In eq.(12), the energy $E$ is proportional to the volume of the earth and inversely proportional to the second law of the distance $r$ approximately. The force is obtained by $F = \partial E/\partial r$. Fig.3 shows the $r$ dependence of the force between an object and the Earth. In Fig.3, the calculated curve is consistent with the relation of $r^{-3}$, and it deviates from the inverse-square law of universal gravitation slightly. In addition, the force F is proportional to the volumes of the earth and object.

The van der Waals force between two atoms decays like $D^{-7}$ and $D^{-8}$ at large $D$ according to eqs.(2)-(4). Therefore, it was thought that the van der Waals force was different from the gravitational force completely. However, eqs.(11) and (12) show that the distance dependence of the van der Waals force between two objects is consistent with the law of universal gravitation approximately. Different from conventional viewpoints, the van der Waals force is a long-range attractive force in macroscopic scale. Based on this result, we argue that the gravitational force is the van der Waals force actually. It means that the gravitation is related to the quantum fluctuations of electron clouds in atoms.

According to eqs.(10)-(12), it is expected that the mass of one object is proportional to its volume and polarizability $N(4\pi\varepsilon_0 a^3)$. When every atom in one neutral object contains more electrons, the object will have a higher polarizability. Then the object tends to produce a higher attractive force. It means that the object has a higher mass.

Eqs.(10) and (11) deviate from the law of universal gravitation slightly. It is still possible to remedy the difference. For example, the van der Waals force out of thermal equilibrium decays with the distance more slowly than the force at equilibrium.[18,19] The surface-atom and surface-surface forces with thermal equilibrium decays like $T/l^4$ and $T/l^3$ at large $l$. The $l$ is the distance between the atom (dielectric half-space) and dielectric half-space. However, the surface-atom force out of thermal equilibrium decays like $(T_2^2 - T_1^2)/l^3$ at large $l$.[18] When the condition $(\varepsilon_2 - \varepsilon_0)/\varepsilon_0 \to 0$ exists in a dielectric half-space, the surface-surface force out of thermal equilibrium decays like $(T_2^2 - T_1^2)/l^2$ at large $l$.[19] The $T_1$ and $T_2$ are the temperatures of the atom (dielectric half-space with $(\varepsilon_2 - \varepsilon_0)/\varepsilon_0 \to 0$) and dielectric half-space. The particular result has been confirmed in experiments.[20] The force F in our model is proportional to the polarizability of bound electrons in atoms. If the result out of thermal equilibrium can be extended to our model, it is expected that the force between two spheres will decays like $(T_2^2 - T_1^2)/D^2$ at large $D$. The relation is consistent with the law of universal gravitation. The temperature effect of the object with lower temperature is ignorable when $T_2 \gg T_1$. It may explain why the temperature effect of one object on the earth is ignorable in the force F.

According to eq.(4), eq.(10) and eq.(11), higher attractive force tends to exist between two objects (or astronomical objects) with higher temperatures. When objects or astronomical objects emit electromagnetic waves to decrease their energies, their temperatures will decrease slowly. Consequently, the gravitation among them will decrease, and the universe tends to spread. In addition, under different conditions (temperature, density and distance) the attractive force among macroscopic objects can be described by different relations in the above model. A single relation on the inverse-square law of universal gravitation is not able to describe all of these details. It may explain why persons have to suppose the existence of dark matter and dark energy in the universe in the past.

In summary, the attraction between macroscopic objects is explained according to van der Waals interaction. In the model, the force with thermal equilibrium deviates from the law of universal gravitation slightly. The gravity on the earth is explained approximately. According to this model, we argue that the van der Waals force and gravitational force are equal to each other essentially. The gravitation in matter is a fluctuation-induced force by long-range correlation. This result may change persons′ fundamental understanding for the gravitation and mass in the future.

Acknowledgment. The work is supported by "985 project" science innovative platform of Shandong University (grant number: 111600819630 25).

Figure Captions

Fig.1. (A) Two nanospheres with separation=5nm. (B) Two nanospheres with separation=1000km. (C) An object on the Earth. (D) Two spheres or two astronomical objects. The $R_1$ and $R_2$ are the radii of the spheres 1 and 2.

Fig.2. (A) Atomic polarizability in a neutral atom. (B) Equivalent circuit of dielectrics. (C) Equivalent circuit of metals. (D) Equivalent circuit of the Earth's core. (E) Equivalent circuit of the Earth's mantle and crust.

Fig.3. Distance dependence of the van der Waals force between the Earth and one object. The open circles are the calculated values of the van der Waals force according to our model. The dashed line shows the relation of $F \propto r^{-3}$. The dotted line shows the relation of $F \propto r^{-2}$.

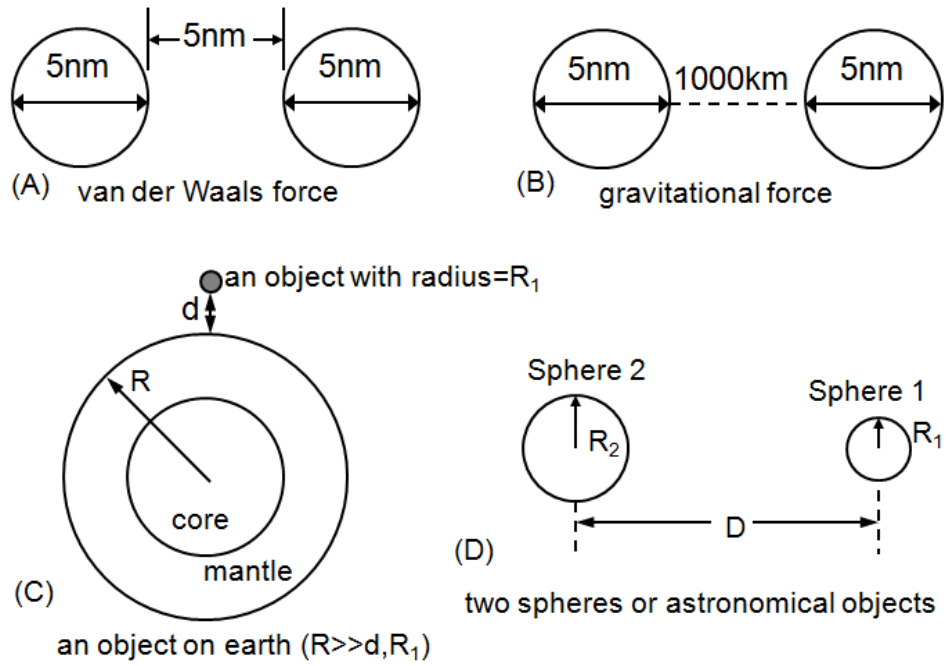

Fig. 1

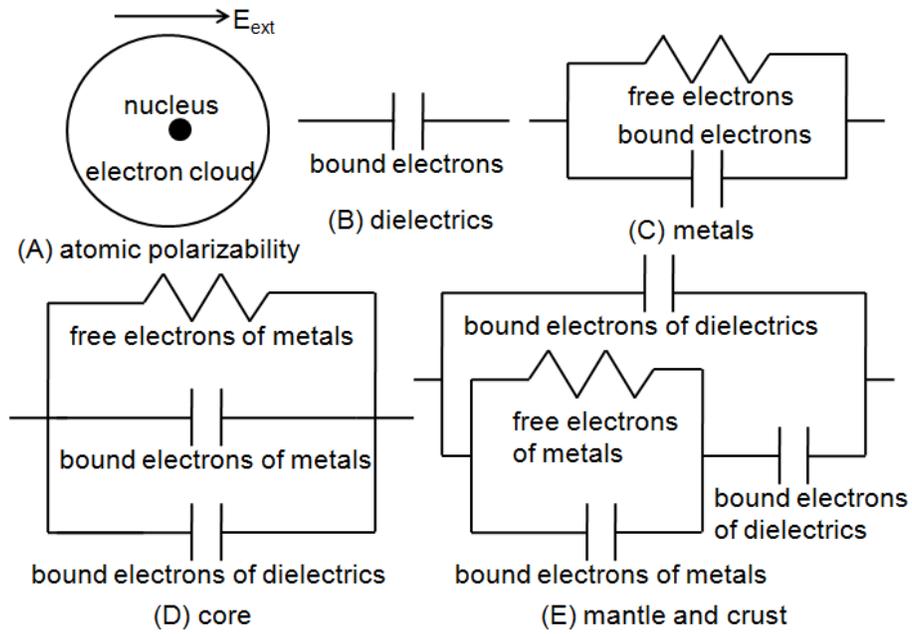

Fig. 2

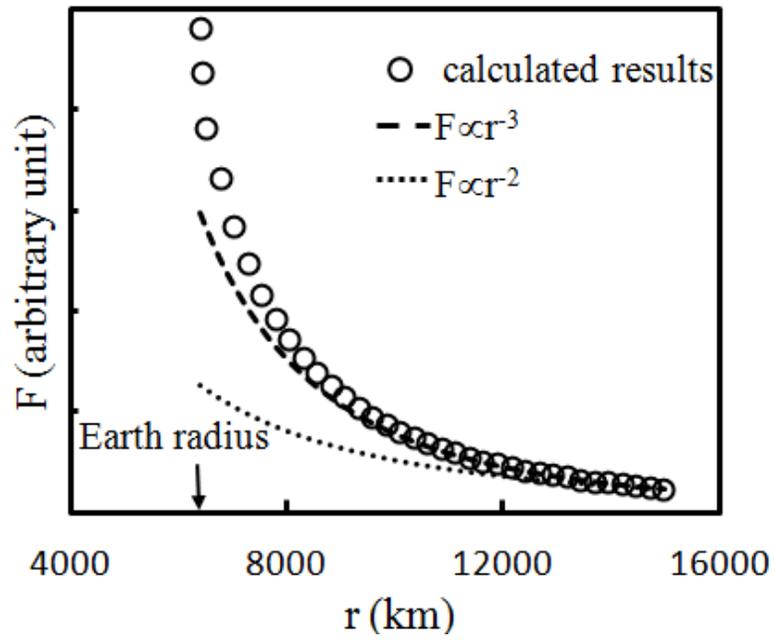

Fig. 3